\documentclass[aps,prx,twocolumn,reprint,amsmath,floatfix,amssymb,showpacs,superscriptaddress]{revtex4-1}
\AtBeginDocument{\usepackage{booktabs}}    
\usepackage{microtype}
\usepackage{wasysym}
\usepackage[varg]{txfonts}
\usepackage[percent]{overpic}
\usepackage{mathrsfs}
\usepackage{braket}
\usepackage{titlesec}
\usepackage{bm}
\usepackage{comment}
\usepackage{verbatim}

\usepackage{array}
\newcommand{\PreserveBackslash}[1]{\let\temp=\\#1\let\\=\temp}
\newcolumntype{C}[1]{>{\PreserveBackslash\centering}p{#1}}
\newcolumntype{R}[1]{>{\PreserveBackslash\raggedleft}p{#1}}
\newcolumntype{L}[1]{>{\PreserveBackslash\raggedright}p{#1}}

\usepackage{color}
\usepackage[colorlinks,bookmarks=false,citecolor=blue,linkcolor=blue,urlcolor=blue,pdfstartview=FitH]{hyperref}

\definecolor{darkred}{rgb}{0.7,0.0,0.0}

\definecolor{darkblue}{rgb}{0,0.02,0.45}

\definecolor{darkgreen}{rgb}{0.02,0.45,0.0}

\definecolor{violet}{rgb}{0.8,0.2,0.6}

\usepackage{amsmath}
\usepackage{amssymb}
\usepackage{graphicx}
\usepackage{subfigure}
\usepackage{pict2e}
\usepackage{multirow}

\newcommand{\be}{\begin{equation}}
\newcommand{\ee}{\end{equation}}
\newcommand{\bea}{\begin{eqnarray}}
\newcommand{\eea}{\end{eqnarray}}
\graphicspath{{Figures/}}

\def\bs{\boldsymbol}
\def\vec{\mathbf}
\def\mc{\mathcal}
\def\nn{\nonumber}

\begin{document}

\title{Collective spin dynamics of $Z_2$ vortex crystals in triangular Kitaev-Heisenberg antiferromagnets}
	
\author{Mengqun Li}
\affiliation{School of Physics and Astronomy, University of Minnesota, Minneapolis, MN 55455, USA}
	
\author{Natalia B. Perkins}
\affiliation{School of Physics and Astronomy, University of Minnesota, Minneapolis, MN 55455, USA}
		
\author{Ioannis Rousochatzakis}
\affiliation{Department of Physics, Loughborough University, Loughborough LE11 3TU, United Kingdom}

\begin{abstract}
We show that the mesoscopic incommensurate $\mathbb{Z}_2$ vortex crystals proposed for layered triangular anisotropic magnets can be most saliently identified by two distinctive signatures in dynamical spin response experiments: The presence of pseudo-Goldstone `phonon' modes at low frequencies $\omega$, associated with the collective vibrations of the vortex cores, and a characteristic multi-scattered intensity profile at higher $\omega$, arising from a large number of Bragg reflections and magnon band gaps. These are direct fingerprints of the large vortex sizes and magnetic unit cells and the solitonic spin profile around the vortex cores.
\end{abstract}
	
\maketitle

\section{Introduction}
\vspace*{-0.35cm}
Recently a significant experimental and theoretical effort has been devoted to the understanding of correlated electron systems with 4d and 5d transition metal ions (like Ru$^{3+}$ and Ir$^{4+}$), characterized by effective $J_{\text{eff}}=1/2$ pseudospins, edge-sharing oxygen octahedra and tri-coordinated lattice geometries~\cite{Jackeli2009,Rau2016,Knolle2017,Winter2017,Trebst2017,Takagi2019}.
Owing to the strong interplay of spin-orbit coupling, crystal field and electronic correlations, these systems show a remarkable range of unconventional phases~\cite{Rau2016,Trebst2017,Knolle2017,Winter2017,Chaloupka2013, Rau2014,Sizyuk2014,Ioannis2015,Gamma2016,Ducatman2018}, including the renowned quantum spin liquids, possibly realized in $\alpha$-RuCl$_3$~\cite{Plumb2014,Sears2015,Johnson2015,Majumder2015,Banerjee2016,Kasahara2018}, the counter-rotating incommensurate spirals realized in the layered honeycomb $\alpha$-Li$_2$IrO$_3$ and its 3D analogues $\beta$- and $\gamma$-Li$_2$IrO$_3$~\cite{Biffin2014a,Biffin2014b,Modic2014,Takayama2015,Veiga2017,Williams2016,Breznay2017,Majumder2018}, and a variety of complex multi-sublattice, single- and multi-${\bf Q}$ phases predicted under a magnetic field~\cite{Janssen2016,Chern2017,Janssen2019}.

The basic ingredient overarching the low-energy descriptions of such systems is the presence of bond-dependent anisotropic exchange, with the so-called Kitaev interactions~\cite{Kitaev2006,Jackeli2009} being the most prominent. As the bond dependence stems from spin-orbit coupling, such interactions are not limited to tri-coordinated lattices, but may also appear in other geometries, including the common frustrated geometries of the triangular, kagome, pyrochlore, and hyperkagome lattices~\cite{Kimchi2014,Jackeli2015,Rousochatzakis2016,Becker2015,Catuneanu2015,Li2015,Shinjo2016,Kos2017,Yao2016,Yao2018,Kishimoto2018}.
In such lattices, the synergy of bond-dependent anisotropy and geometric frustration opens up the possibility for novel cooperative phases even when the anisotropy is not the dominant interaction, as in the above tri-coordinated systems.

\begin{figure*}[!]
\includegraphics[width=\textwidth,trim= 0 -10 0 -10,clip]{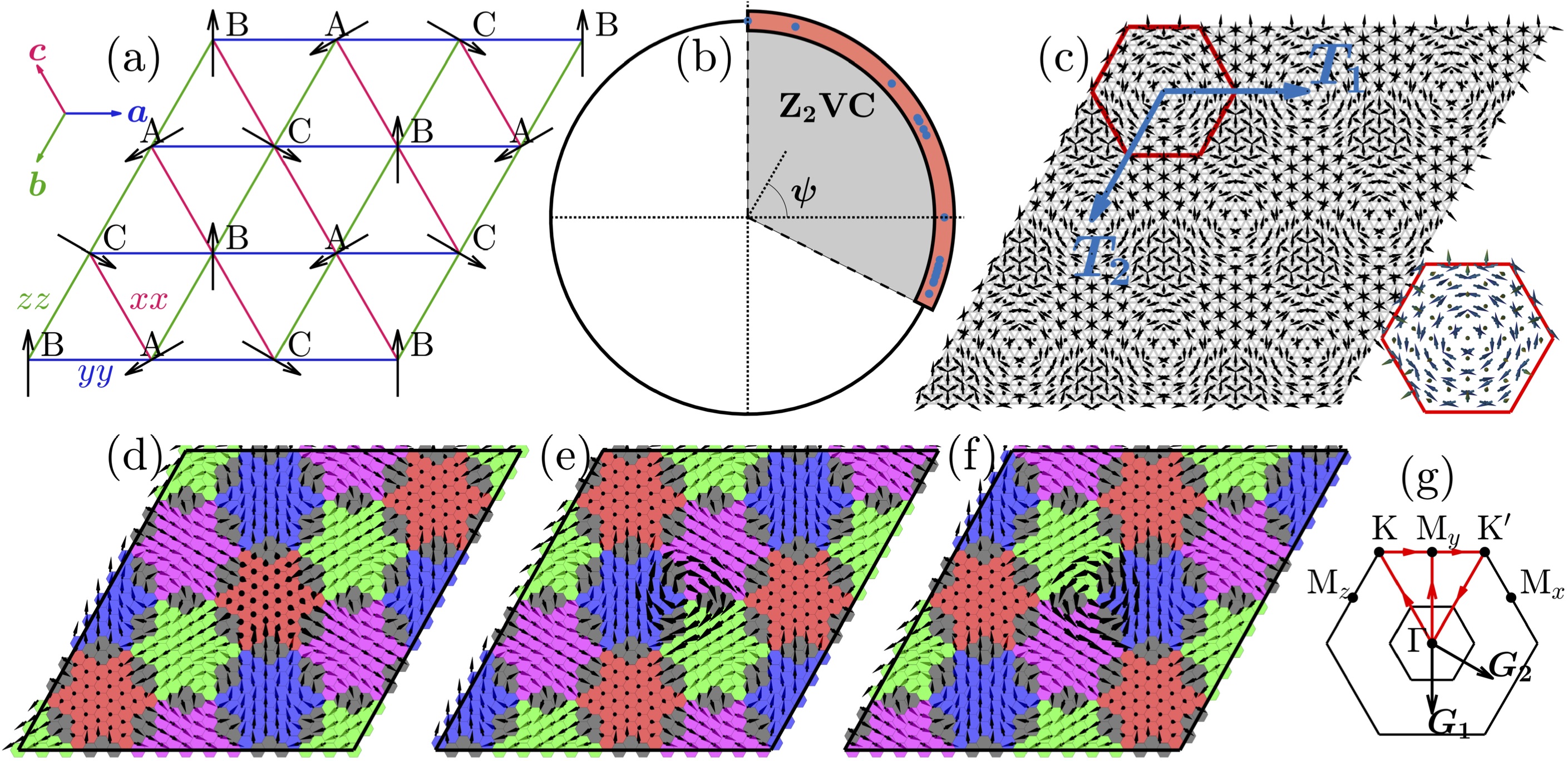}
\caption{
(a) Triangular lattice with three types of bonds, `xx' (red), `yy' (blue) and `zz' (green), along with the 120$^\circ$ order with the three sublattices, $A$, $B$ and $C$. 
(b) Parameter space of the $JK$ model,  along with the stability region (shaded) of the $\mathbb{Z}_2$VC phase, $\psi\!\in\!(\tan^{-1}(-1/2)$, $\pi/2)$. Blue dots depict the parameter points (and corresponding optimal crystals) analyzed in this study. 
(c) The spin pattern of the optimal $\mathbb{Z}_2$VC stabilized at $\psi\!=\!\tan^{-1}(-0.34)$, for which $d\!=\!7$, $\mc{N}_m\!=\!196$. $\vec{T}_1$ and $\vec{T}_2$ are the translation vectors of the superlattice. (d-f) Separate spin patterns on sublattices $A$, $B$ and $C$. The color indicates the dominant spin projection onto one of the four $\langle111\rangle$ symmetry axes ($[111]$ (red), $[1\bar{1}\bar{1}]$ (green), $[\bar{1}1\bar{1}]$ (blue), $[\bar{1}\bar{1}1]$ (magenta)), and onto one of the three $\langle100\rangle$ axes (grey). The spin texture associated with one vortex is highlighted by bold arrows. (g) Lattice Brillouin zone (BZ, outer hexagon) and magnetic BZ (inner hexagon, not in scale).}
\label{fig:lattice}
\end{figure*}

Already the introduction of an infinitesimal Kitaev anisotropy $K$ in one of the simplest frustrated geometries, the triangular lattice [Fig.~\ref{fig:lattice}\,(a)], highlights the prolificacy of this synergy~\cite{Rousochatzakis2016}: The three-sublattice 120$^\circ$ order of the triangular Heisenberg antiferromagnet (HAF) is immediately unstable under $K$, giving way to incommensurate crystals of $\mathbb{Z}_2$ vortices of {\it mesoscopic} size [Figs.~\ref{fig:lattice}\,(c-f)], see also [\onlinecite{Becker2015,Catuneanu2015,Shinjo2016,Trebst2017,Li2015,Kos2017,Kishimoto2018}]. 
Such vortices have been known~\cite{Kawamura:1984hk} to be present in triangular HAFs as topological excitations, but here the bond-dependent anisotropy condenses them in the ground state via a commensurate-incommensurate (C-IC) nucleation mechanism~\cite{DesGennes1975,Bak1982,McMillan1976,Schaub1985}. This is akin to the formation of magnetic domains~\cite{DesGennes1975}, Abrikosov vortices~\cite{Abrikosov1957,Essmann1967,Hess1989}, blue phases in cholesteric liquid crystals~\cite{Wright89}, skyrmions in chiral helimagnets~ \cite{Bogdanov1989,Roessler2006,muehlbauer2009,tonomura2012,yu2012,yu2010,Seki2012,oleg2014}, and other systems~\cite{Suzuki1983,Okubo2012,Leonov2015,Rosales2015,Hayami2016}. 
Anisotropic antiferromagnets with hexagonal symmetry provide, therefore, a fertile ground for novel incommensurate phases with topological, particle-like properties.

While the prospect of realizing the $\mathbb{Z}_2$ vortex phase remains currently open (see Sec.~\ref{sec:Discussion} below), here we explore the collective spin dynamics in this phase and demonstrate numerically how its presence can be most saliently observed in dynamical spectroscopic probes. 
To this end, we construct a large family of $\mathbb{Z}_2$ vortex crystals ($\mathbb{Z}_2$VC's), for both positive and negative Kitaev anisotropy -- with magnetic unit cell sizes extending up to $2028$ spin sites -- and perform a semiclassical $1/S$ expansion to extract the magnon spectrum, the associated spin dynamical structure factors (DSF) $\mc{S}^{\alpha\beta}({\bf Q},\omega)$ (for all relative polarizations $\alpha, \beta=\{x,y,z\}$), and the corresponding inelastic neutron scattering (INS) intensity $\mc{I}({\bf Q},\omega)$.

The results close to the C-IC transition mirror two of the most distinctive features of the $\mathbb{Z}_2$VC phase, the large size of the vortices and their particle-like nature. Conceptually, both of these features stem from the C-IC nature of the transition from the `parent' 120$^\circ$ state~\cite{Rousochatzakis2016}. The vortex size is large close to the transition because the vortices play the role of `discommensurations' of the parent state, and their relative distance must diverge when we recover that state. This manifests in $\mc{I}({\bf Q},\omega)$ by a distinguished multi-fragmentation of the `parent' magnon bands, arising from a high density of Bragg reflections.

\begin{figure}[!b]
\centering
\includegraphics[width=0.99\columnwidth,trim= 0 -10 0 -10,clip]{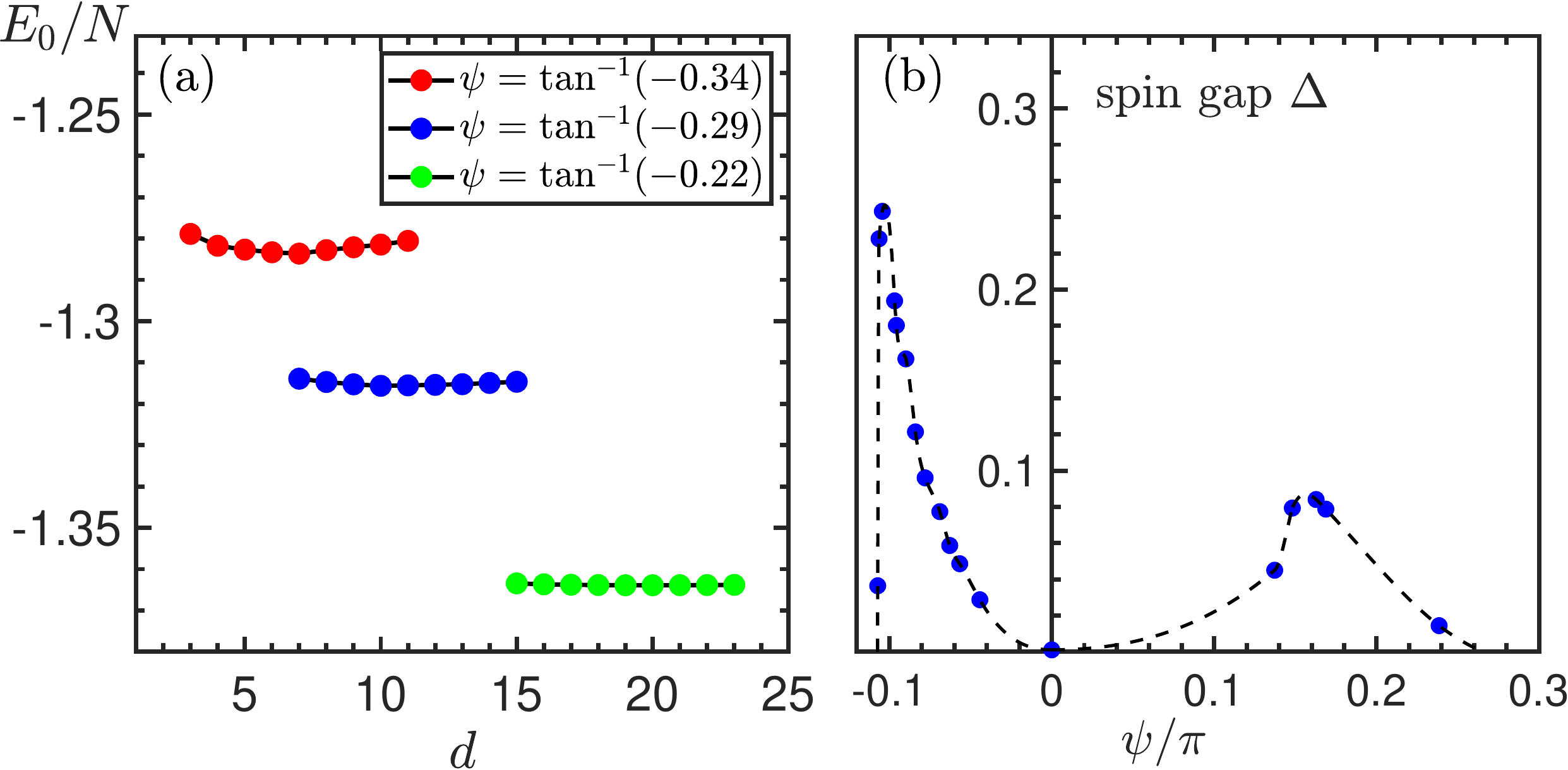}
\caption{(a) Energy per site vs $d$ for $K/J\!=\!-0.34$ (red), $-0.29$ (blue) and $-0.22$  (green). (b) Evolution of the spin gap $\Delta$ with $\psi$.}
\label{fig:Ed}
\end{figure}

The particle-like character of the vortices manifests at low frequencies $\omega$ via the presence of intense pseudo-Goldstone modes. These modes are associated with collective vibrations of the vortex cores around their equilibrium positions, and are thus analogous to phonons in crystals. Their appearance attests to the {\it nonlinear} character of the spin profile around the cores. As shown in [\onlinecite{Rousochatzakis2016}], the vortices arise by a special intertwining of three honeycomb superstructures of ferromagnetic (FM) domains [one for each sublattice of the `parent' 120$^\circ$ phase, see Fig.~\ref{fig:lattice}\,(d-f)], and this arrangement gives rise to abrupt, soliton-like modulations around the vortex cores.  
As demonstrated below [Fig.~\ref{fig:Ed}\,(a)], the ground state energy landscape (as a function of the core positions) flattens significantly as we approach the C-IC transition,
revealing a weak inter-particle potential at large distances. 
The pseudo-Goldstone modes (which are otherwise gapped out by the lattice cutoff) are thus a manifestation of the nonlinear spin profiles of the cores. 

The paper is organized as follows. We begin with the definition of the model (Sec.~\ref{sec:Model}), a brief review of the main features of the $\mathbb{Z}_2$VC's (Sec.~\ref{sec:Z2VC}), and the iterative variational method used to obtain optimal crystals for given model parameters (Sec.~\ref{sec:VM}). Our results for the collective spin dynamics and the corresponding predictions for the inelastic neutron scattering intensity are presented in Sec.~\ref{sec:DF}. A brief outlook is given in Sec.~\ref{sec:Discussion}, while auxiliary information and technical details are relegated to Apps.~\ref{app:MUCs}-\ref{app:RepINS}.

\vspace*{-0.15cm}
\section{Model}\label{sec:Model}
\vspace*{-0.35cm}
The Hamiltonian of the Heisenberg-Kitaev or $JK$-model~\cite{Jackeli2009} on the triangular lattice reads
\begin{align}\label{eq:ham}
\mc{H} = \sum\nolimits_{\langle ij\rangle} \left( J ~\vec{S}_i\!\cdot\!\vec{S}_j + K ~S_i^{\gamma_{ij}}S_j^{\gamma_{ij}} \right)\,.
\end{align}
Here $\langle ij\rangle$ denotes nearest neighbor lattice sites, ${\bf S}_i$ and ${\bf S}_j$ are pseudospins-1/2 degrees of freedom, and $J$ and $K$ denote the Heisenberg and Kitaev exchange parameter, respectively. The component $\gamma_{ij}$ is given by 
\be
\gamma_{ij}\!=\!x,~ y,~ \text{or}~ z, 
\ee
depending on whether $\langle ij\rangle$ belongs to the `xx', `yy' or `zz' type of bonds, see Fig.~\ref{fig:lattice}\,(a). The lattice plane is (111), and the vectors ${\bf a}$, ${\bf b}$ and ${\bf c}$ shown in Fig.~\ref{fig:lattice}\,(a) point along ${\bf z}\!-\!{\bf x}$,  ${\bf x}\!-\!{\bf y}$ and ${\bf y}\!-\!{\bf z}$, respectively. 
In what follows we use the parametrization $J\!=\!\cos\psi$ and $K\!=\!\sin\psi$ and restrict ourselves to the stability region $\psi\!\in\!(\tan^{-1}(-1/2)$, $\pi/2)$ of the vortex phase  [shaded in Fig.~\ref{fig:lattice}\,(b)]~\cite{Rousochatzakis2016,Becker2015}. We also set the lattice parameter $a\!=\!1$.

The ground state of the HAF point ($\psi\!=\!0$) of the $JK$ model (Eq.~(\ref{eq:ham})) is the well-known 120$^\circ$, three-sublattice coplanar order~\cite{Yafet1952}, whose order parameter is that of a rigid rotator, i.e., SO(3). 
Classical analysis~\cite{Rousochatzakis2016} shows that the 120$^\circ$ pattern is immediately unstable under an infinitesimal Kitaev interaction, giving way to a nontrivial, long-distance twisting of the SO(3) order parameter in both directions of the lattice plane, leading to localized $\mathbb {Z}_2$ vortices (see also \cite{Becker2015,Trebst2017}). 
The cores of the vortices form a triangular superstructure whose period $d$ (the distance between the vortex cores) is determined by the competition between the Kitaev exchange $K$ and the Heisenberg exchange $J$. For small $|K|/J$, $d\propto J/|K|$, i.e., the distance between vortex cores goes to infinity at the HAF point, and the transition between the 120$^\circ$ order and the $\mathbb {Z}_2$VC phase is of the C-IC nucleation type~\cite{DesGennes1975,Bak1982,McMillan1976,Schaub1985}.

\vspace*{-0.15cm}
\section{Main aspects of the $\mathbb{Z}_2$VC phase}\label{sec:Z2VC}
\vspace*{-0.35cm}
Let us recall the main features of the $\mathbb{Z}_2$VC phase~\cite{Rousochatzakis2016}. First, the cores of the vortices are defects of the $120^{\circ}$ state, as they are associated with a finite FM canting and a reduced vector chirality. This means that the cores cost Heisenberg energy. However, the Kitaev energy around the cores is negative, which is why having cores with a given density is energetically favorable.

Second, the distance $d$ between the cores (Fig.~\ref{fig:lattice}\,(c)) and the size of the vortices are infinite at $\psi\!=\!0$ (HAF point), and decreases monotonously as we depart from this point. The minimum values of $d$ are reached at the phase boundaries with the neighboring phases at $\tan\psi\!=\!-1/2$ ($d\!=\!1$) and $\psi\!=\!\pi/2$ ($d\!=\!2$). 
The magnetic unit cell contains $\mc{N}_m\!=\!4d^2$ or $\mc{N}_m\!=\!12 d^2$ spins, depending, respectively, on whether the translation vectors of the state, $\vec{T}_1$ and $\vec{T}_2$ [see Fig.~\ref{fig:lattice}\,(c)], map spins from one type of sublattice to another or not~\footnote{The second possibility has been overlooked in Ref.~[\onlinecite{Rousochatzakis2016}].}, see detailed discussion in Sec.~\ref{sec:VM} and Table~\ref{tab:d_req_1}.

Third, the anatomy of the $\mathbb{Z}_2$VC can be best understood by visualizing separately the spins in the three sublattices of the 120$^\circ$ state, see Figs.~\ref{fig:lattice}\,(d-f). In contrast to the 120$^\circ$ state, where all spins of a given sublattice are parallel to each other, forming a single FM domain of infinite size, here the spins of a given sublattice form a hexagonal superstructure of FM domains. 
The $\mathbb{Z}_2$ vortices then arise by the special way the three sublattice superstructures are intertwined with each other. In particular, the center of a FM domain in one sublattice (say $A$) coincides with vertices of the hexagonal superstructures in the other two sublattices ($B$ and $C$). 
Therefore, as we trace a closed loop around the center of a FM domain of $A$, the spins of $A$ remain roughly parallel along the loop, whereas the spins of $B$ and $C$ complete a $2\pi$-rotation, leading to a $\mathbb{Z}_2$ vortex, see bold arrows in Fig.~\ref{fig:lattice}\,(e-f).
The precise way the $2\pi$-rotation happens is related to the special role of the $[111]$ and $[100]$ directions, see color coding of Figs.~\ref{fig:lattice}\,(d-f) and detailed discussion in Ref.~[\onlinecite{Rousochatzakis2016}]. 

\begin{figure*}[!t]
\centering
\subfigure[$\tan\psi\!=\!0.502$, ~$d\!=\!11$, ~$\mc{N}_m\!=\!484$]{\includegraphics[width=0.495\textwidth,trim= 0 -10 0 0,clip]{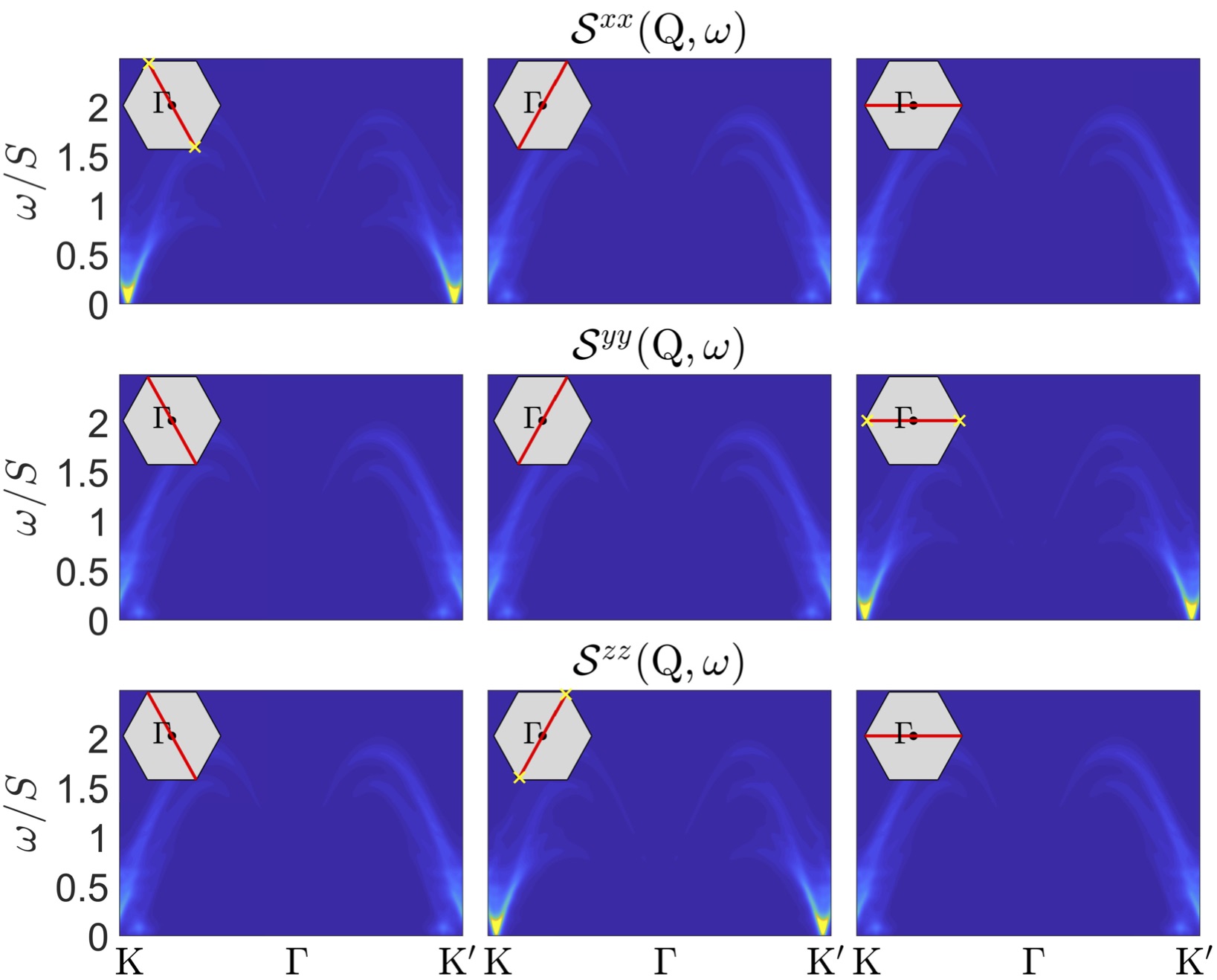}}
\subfigure[$\tan\psi\!=\!-0.34$, ~$d\!=\!7$, ~$\mc{N}_m\!=\!196$]{\includegraphics[width=0.495\textwidth,trim= 0 0 0 -10,clip]{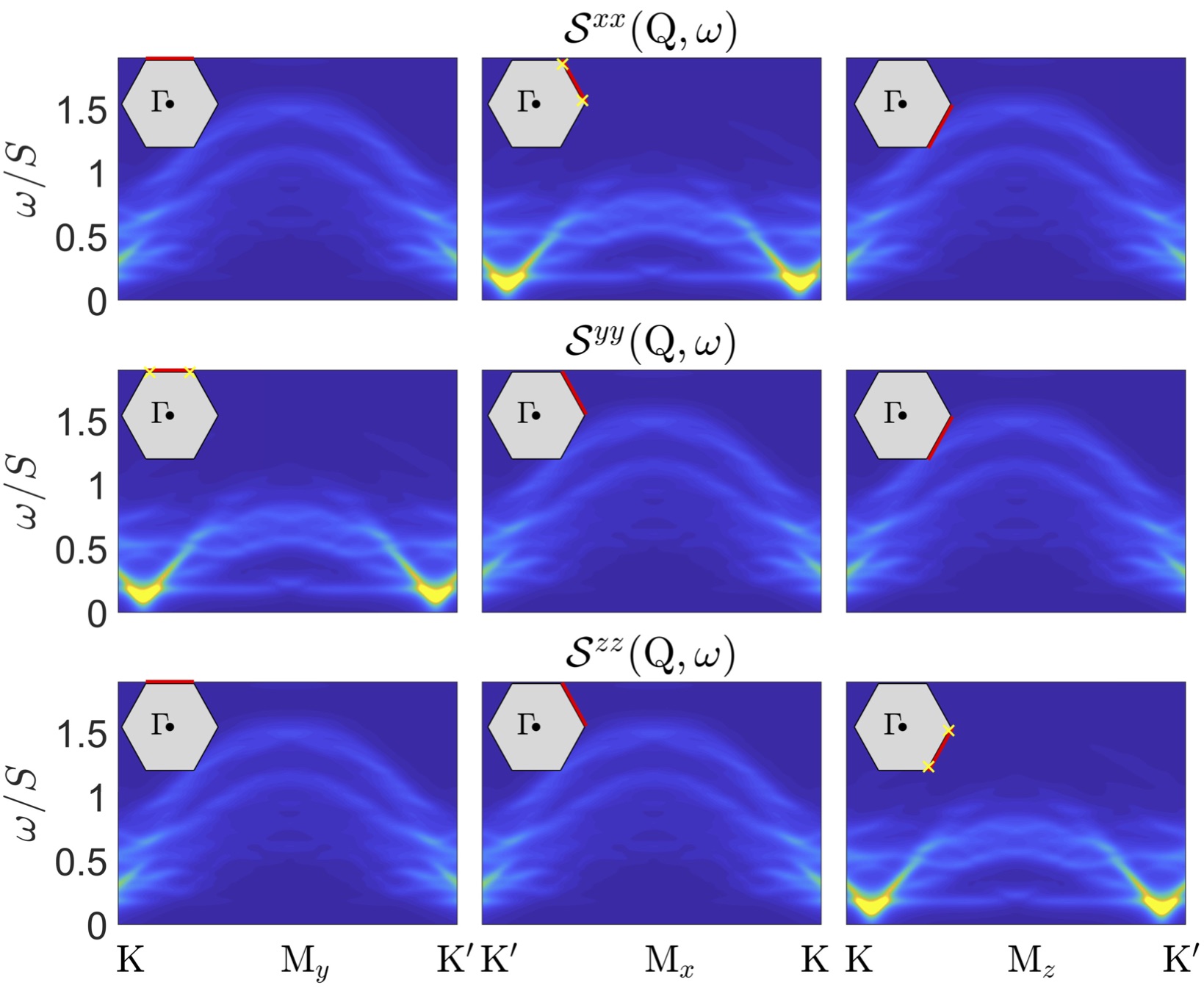}}
\caption{DSF ${\mathcal S}^{xx}(\vec{Q},\omega)$, ${\mathcal S}^{yy}(\vec{Q},\omega)$, ${\mathcal S}^{zz}(\vec{Q},\omega)$ computed for (a) $\tan\psi\!=\!0.502$ and (b) $\tan\psi\!=\!-0.34$ along $\mathrm{K}\rightarrow\mathrm{\Gamma}\rightarrow\mathrm{K}'$ and $\mathrm{K}\rightarrow\mathrm{M}\rightarrow\mathrm{K}'$ high symmetry paths (insets, red lines), respectively. Yellow crosses show the positions of the first harmonic peaks of the static spin structure factor. 
The color scale runs from ‘blue’ corresponding to the minimum intensity to ‘red’ corresponding to the maximum intensity, and it is independently normalized for each plot.
}
\label{fig:DSF}
\end{figure*}

Finally, the $\mathbb{Z}_2$VC state preserves the discrete threefold rotation symmetry of the model. As we show below, this gives rise to three pairs of pseudo-Goldstone modes which are related to each other by threefold rotations. These modes track  the first harmonic Bragg peaks in the static spin structure factor~\cite{Rousochatzakis2016,Becker2015}. Namely, they emanate from the corners of the BZ as we depart from the HAF point, and move towards the $\Gamma$ point for $K\!>\!0$, or the ${\bf M}$ points for $K\!<\!0$ [see Figs.~\ref{fig:DSF}-\ref{fig:scattering} below].

\vspace*{-0.15cm}
\section{Optimal crystals and variational minimization method}\label{sec:VM}	
\vspace*{-0.35cm}
For each given $\psi$ inside the stability region of the $\mathbb{Z}_2$VC phase, the optimal value of $d$ can be obtained by the variational energy minimization scheme outlined in [\onlinecite{Rousochatzakis2016}]. In this approach, one exploits the fact that the $\mathbb {Z}_2$VC's consist of three honeycomb superstructures of ferromagnetic domains, one for each of the three sublattices (A, B and C) of the HAF point. The majority of spins within each FM domain point along a specific direction in spin space, which happens to be one of the four $\langle 111\rangle$ axes. We therefore begin by constructing, for each given $\psi$, an initial state consisting of perfect FM domains (where all spins in the domain are parallel to each other and along the respective $\langle 111\rangle$ axis) with a size that corresponds to a fixed choice of $d$. Next, upon a random sampling, we sequentially rotate spins in the direction of their local mean fields. After a certain number of samplings, the system converges to a $\mathbb{Z}_2$VC and the corresponding energy per site $E_0(d)/N$ is extracted. This procedure is repeated for a series of different FM domain wall sizes, corresponding to different choices of $d$ (and always using appropriate clusters with periodic boundary conditions that accommodate the given superstructure). The resulting energies per site $E_0(d)/N$ are then plotted as a function of $d$ and one identifies the optimal crystal with the one associated with the minimum energy. Three examples were shown in Fig.~\ref{fig:Ed}\,(a), for $\tan\psi\!=\!-0.34$, $-0.29$ and $-0.22$, for which the minimum energies per site are reached at $d\!=\!7$, $10$ and $19$, respectively. 
Following these steps we construct a large set of optimal crystals [see blue dots in Fig.~\ref{fig:lattice}\,(b)], with $d$ extending from $1$ ($\tan\psi\!=\!-0.42$, $\mc{N}_m\!=\!4$ spins in the magnetic unit cell) to $d\!=\!19$ ($\tan\psi\!=\!-0.22$, $\mc{N}_m\!=\!1444$) for negative $K$, and from $d\!=\!2$ ($\tan\psi\!=\!4$, $\mc{N}_m\!=\!16$) to $d\!=\!13$ ($\tan\psi\!=\!0.46$, $\mc{N}_m\!=\!2028$) for positive $K$.

\begin{figure*}[!t] 
\includegraphics[width=\textwidth,trim= 0 0 0 0,clip]{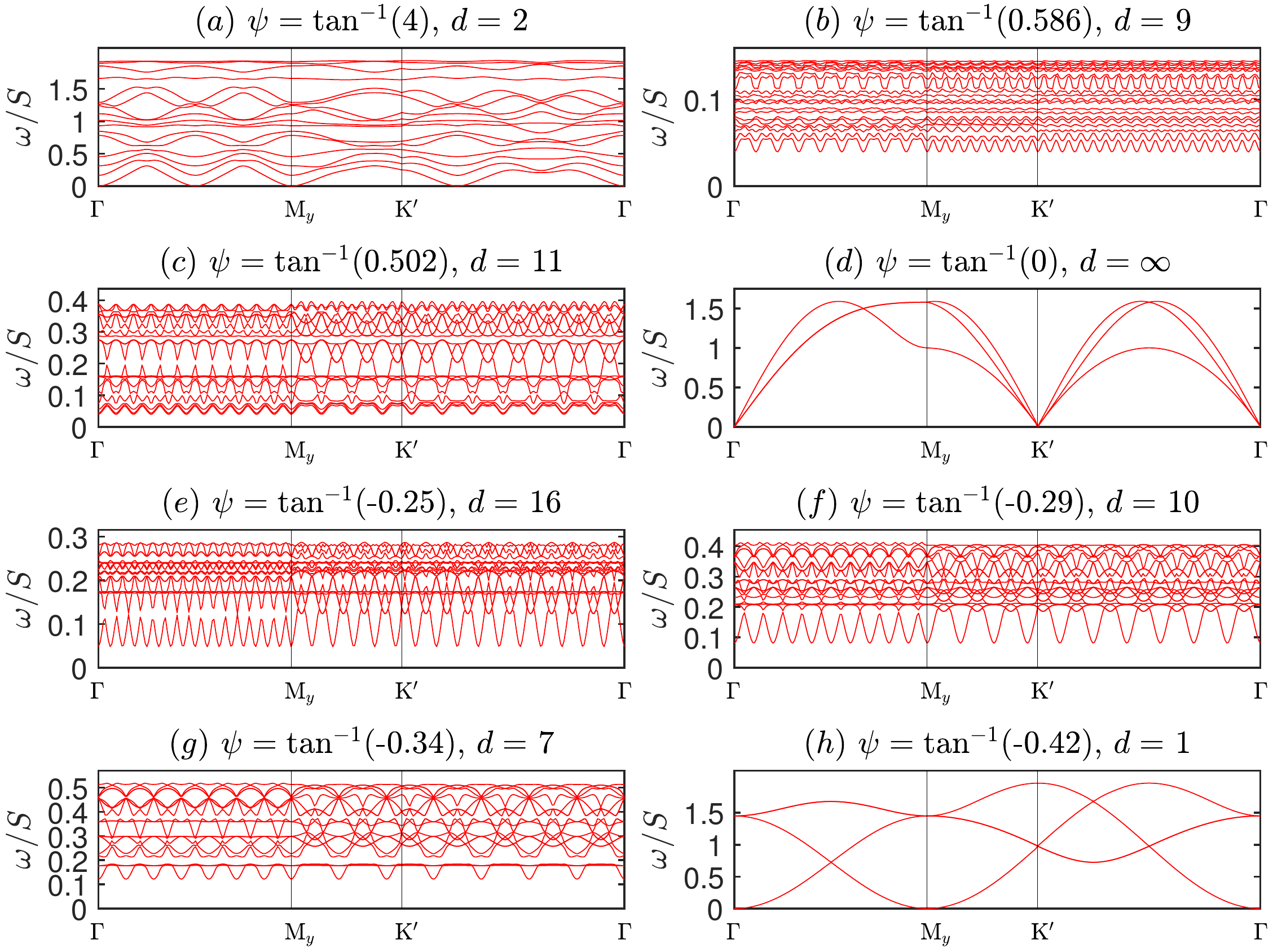}
\caption{Linear spin wave branches along the symmetry path $\mathrm{\Gamma} \rightarrow \mathrm{M}_y \rightarrow \mathrm{K}' \rightarrow \mathrm{\Gamma}$ of the lattice BZ, computed for $\tan\psi\!=\!4$ (a), $0.586$ (b), $0.502$ (c), $0$ (d), $-0.25$ (e), $-0.29$ (10), $-0.34$ (g) and $-0.42$ (h). Only the lowest 20 branches are shown here when $\mc{N}_m>20$.}
\label{fig:spinwave}
\end{figure*}

\vspace*{-0.15cm}
\section{Dynamical fingerprints of the $\mathbb{Z}_2$VC phase}\label{sec:DF}
\vspace*{-0.35cm}
The collective spin dynamics  can now be studied, for each of these optimal crystals, using a numerical implementation of the Holstein-Primakoff transformation, followed by a generalized Bogoliubov transformation, and a numerical diagonalization that delivers the $\mc{N}_m$ magnon bands in the magnetic BZ. This is then used for the evaluation of 
\be
\mc{S}^{\alpha\beta}({\bf Q},\omega)\!=\!\int\text{dt}\,e^{-i\omega t}\langle S^\alpha(-\vec{Q},0)S^\beta(\vec{Q},t)\rangle\,,
\ee 
where ${\bf S}(\vec{Q},t)$ is the Fourier transform of the total spin with ${\bf Q}$ in the first BZ of the lattice,  and 
\be
\mc{I}({\bf Q},\omega)\!\propto\!\sum_{\alpha\beta}(\delta_{\alpha\beta}-Q^\alpha Q^\beta/Q^2)~\mc{S}^{\alpha\beta}(\vec{Q},\omega)\,,
\ee 
for further technical details see App.~\ref{app:LSWS}.

\vspace*{-0.15cm}
\subsection{Linear spin wave (LSW) expansion}
\vspace*{-0.35cm}
Figure~\ref{fig:spinwave} shows the LSW dispersions for eight representative optimal $\mathbb{Z}_2$VC's. The spectra are first obtained in the magnetic BZ and then plotted in the repeated scheme, along special symmetry directions in the lattice Brillouin zone (see hexagons in Fig.~\ref{fig:translation}).  
Panel (d) shows the familiar result for the 120$^{\circ}$-order of the pure HAF ($\psi\!=0$), which can actually be considered as a $\mathbb{Z}_2$VC state with $d=\infty$~\cite{Mourigal2013}.
As we gradually move away from the HAF point, the size of the $\mathbb{Z}_2$-vortex becomes finite but still remains very large. Recall that the size of magnetic unit cell is $\mc{N}_m=4 d^2$ or $12 d^2$, depending on the orientation of the spanning vectors of the superlattice, see above. This explains the large number of $\mc{N}_m$ magnon bands that are visible in Fig.~\ref{fig:spinwave}, except for panels (d) and (h). 
The figure also shows the band gaps between neighboring magnon bands, which result from Bragg reflections of the spin waves off the boundaries of the large magnetic unit cells. 
This high density of Bragg reflections and magnon band gaps is responsible for the multi-fragmented scattering profile announced above.

\vspace*{-0.15cm}
\subsection{Spin dynamical structure factors (DSF)}
\begin{figure*}[!t]
\includegraphics[width=0.99\textwidth,trim= 0 0 0 0,clip]{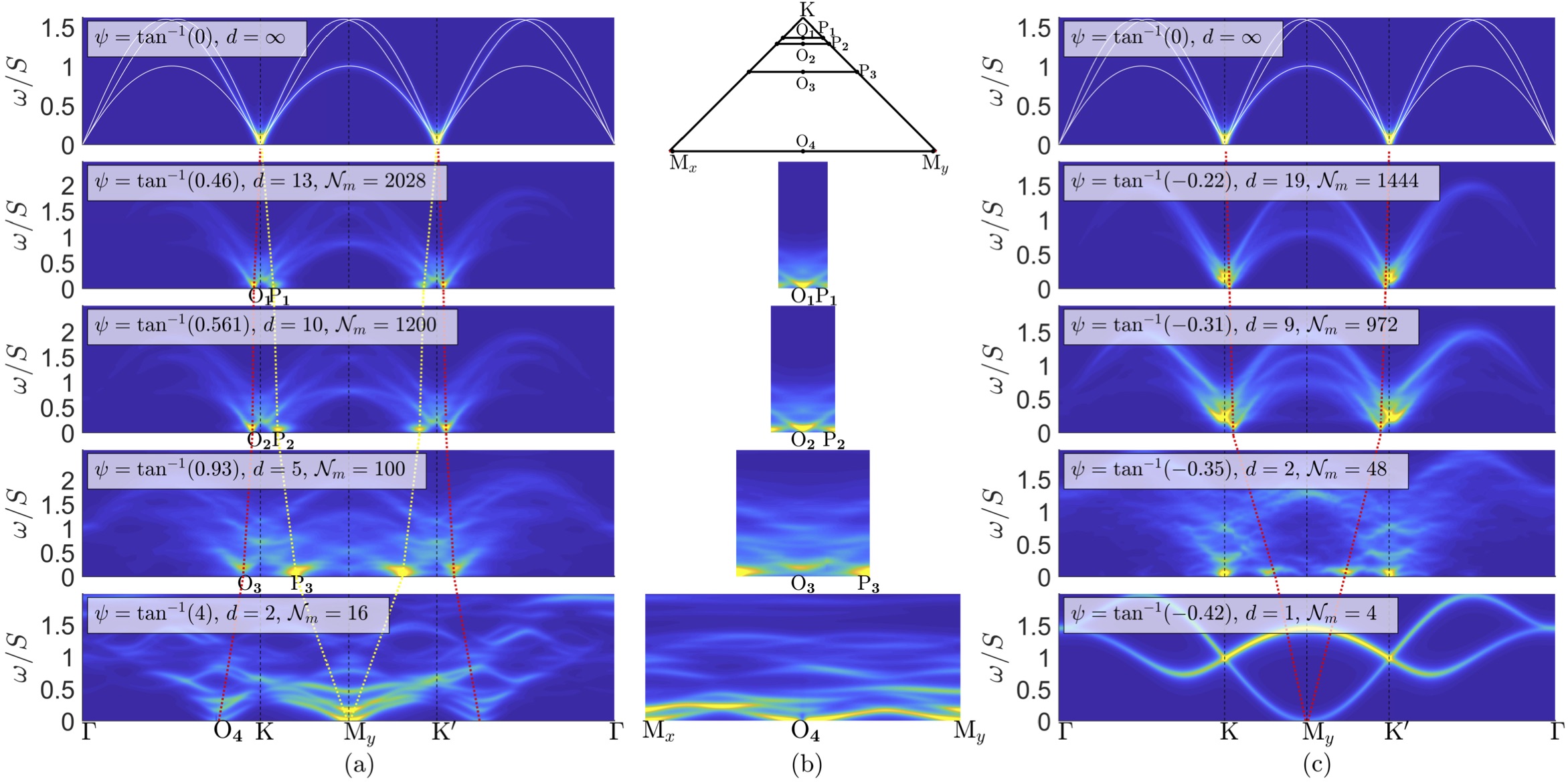}
\caption{INS intensities along a high symmetry path in the first BZ ($\mathrm{\Gamma} \rightarrow \mathrm{K} \rightarrow \mathrm{M}_y \rightarrow \mathrm{K}' \rightarrow \mathrm{\Gamma}$) defined in Fig.~\ref{fig:lattice}\,(g) for eight representative $\mathbb{Z}_2$VC states, where $K\geqslant0$ in panel (a) and $K\leqslant0$ in panel (c). Shifting of the minima is marked by the red dotted lines. The dispersion at HAF point is also given for comparison. (b) INS intensities along the line where these minima $\mathrm{O_i}$, $\mathrm{P_i}$ ($\mathrm{P_4}$ coincides with $\mathrm{M}_y$) in (a) are residing. The relative position of these lines in the BZ are shown in the top panel of (b).} 
\label{fig:scattering}
\end{figure*}

\vspace*{-0.35cm}	
Figures~\ref{fig:DSF}\,(a-b) show the diagonal components ${\mc S}^{xx}(\vec{Q},\omega)$, ${\mc S}^{yy}(\vec{Q},\omega)$ and ${\mc S}^{zz}(\vec{Q},\omega)$ for two representative optimal crystals with large $d$, one at $\tan\psi\!=\!0.502$ ($d\!=\!11$, $\mc{N}_m\!=\!484$) and the other at $\tan\psi\!=\!-0.34$ ($d\!=\!7$, $\mc{N}_m\!=\!196$). 
First of all, it can be clearly seen that the three diagonal components are indeed related to each other by the threefold symmetry. 
Second, the overall shape of the DSF at intermediate and high $\omega$ (i.e., far enough from the corners of the lattice BZ) follows very roughly the shape of the three magnon bands of the DSF of the parent 120$^\circ$ order (see Fig.~\ref{fig:spinwave}\,(d) and top panels in Fig.~\ref{fig:scattering}, as well as Ref.~\cite{Mourigal2013}). Equivalently, the unfolded (in the lattice BZ) $\mc{N}_m$ magnon bands of the $\mathbb{Z}_2$VC follow roughly the overall shape of the three `parent bands'. This is due to the fact that the magnon wavelengths in this part of the spectrum can be significantly smaller than the distance $d$ between vortices, and the short-distance fluctuations are still governed by the Heisenberg exchange. 
Despite this rough similarity, the huge number of Bragg reflections and associated band gaps (resulting from the large magnetic unit cell) give rise to a qualitative different DSF,  with only a small portion of the bands standing out and an otherwise smeared out and multi-fragmented response.

The most intense modes in Figs.~\ref{fig:DSF}\,(a-b) appear at low $\omega$, close to the corners of the BZ, where the magnon wavelengths become comparable to the distance $d$ between vortices. These intense modes are the collective, pseudo-Goldstone modes mentioned above, associated with the rigid vibrations of the vortex cores around their equilibrium positions. There are three ($\pm{\bf Q}$) pairs of such phonon-like modes [one for each diagonal component of $\mc{S}^{\alpha\alpha}({\bf Q},\omega)$], and their positions coincide with those of the first harmonics of the static structure factor~\cite{Rousochatzakis2016,Becker2015}, see yellow stars in the insets of Fig.~\ref{fig:DSF}. 
All in all, Fig.~\ref{fig:DSF} therefore demonstrates the two most salient dynamical fingerprints of the $\mathbb{Z}_2$VC phase in the vicinity of the C-IC transition, the large vortex size and their particle-like character.

\vspace*{-0.15cm}
\subsection{Evolution of the spectra with $K/J$}\label{sec:Evolution}
\vspace*{-0.35cm}
We now proceed to elucidate the way these features evolve as we move deeper into the $\mathbb{Z}_2$VC phase, and the vortices become smaller in size. To this end, we consider a series of eight representative $\mathbb{Z}_2$VC's with decreasing $d$, four for $K\!>\!0$ [panels (a-b), $d\!=\!\{13, 10, 5, 2\}$] and four for $K\!<\!0$ [panel (c), $d\!=\!\{19, 9, 2, 1\}$]; For results on many more representative states see App.~\ref{app:RepINS}.  
Figure~\ref{fig:scattering} shows the associated $\mc{I}(\vec{Q},\omega)$, along with the intensity of the HAF point (top panels, $d\!=\!\infty$).  
The rough resemblance mentioned above, between the overall shape of the response with that of the parent state, persists down to $d\!=\!5$ and $\mc{N}_m\!=\!100$ for $K\!>\!0$, and down to $d\!=\!2$ and $\mc{N}_m\!=\!48$ for $K\!<\!0$. For smaller vortex sizes new features appear, such as the distinctively rich pattern for $\tan\psi\!=\!4$ (panel a, bottom) which is characteristic of strong Kitaev physics (see also below), and the two-band picture for $\tan\psi\!=\!-0.42$, which is characteristic of the neighboring $\widetilde{\text{FM}}$ state~\cite{Rousochatzakis2016,Becker2015}. 

Turning to the evolution of the phonon-like modes, these must track the positions of the first harmonic Bragg peaks, as mentioned above. This is illustrated by the red dashed lines in panels (a) and (c). For $K\!>\!0$ (panel a), one of the phonon modes traces the path $\mathrm{K}\!\to\!\mathrm{O}_1\!\to\!\mathrm{O}_2\!\to\!\mathrm{O}_3\!\to\!\mathrm{O}_4$, while for $K\!<\!0$ (panel c) the phonon mode shown goes from $\mathrm{K}\to\mathrm{M}_y$, and similarly for the remaining phonon modes related by threefold rotations.

In addition to the phonon modes, we also find a second intense low-$\omega$ mode. For $K\!>\!0$, this is shown by the dashed yellow line in panel (a) and is elucidated further in panel (b).  This mode traces the path $\mathrm{K}\!\to\!\mathrm{P}_1\!\to\!\mathrm{P}_2\!\to\!\mathrm{P}_3\!\to\!\mathrm{M}_y$, and is a precursor of an accidental, classical ground state degeneracy present at the Kitaev point ($\psi\!=\!\pi/2$)~\cite{Rousochatzakis2016}. This degeneracy is sub-extensive and manifests in the Fourier transform of the classical energy with lines of minima joining the ${\bf M}$ points of the BZ (e.g., the line $\mathrm{M}_x\to\mathrm{M}_y$). This is illustrated in panel (b) which shows the intensity along special horizontal cuts (panel b, top) parallel to $\mathrm{O}_1\mathrm{P}_1$ ($\tan\psi\!=\!0.46$), $\mathrm{O}_2\mathrm{P}_2$ ($\tan\psi\!=\!0.561$), $\mathrm{O}_3\mathrm{P}_3$ ($\tan\psi\!=\!0.93$) and $\mathrm{O}_4\to\mathrm{M}_y$ ($\tan\psi\!=\!4$). The intensities along these cuts show the development of an almost flat mode, which should ideally become completely flat at $\psi\!=\!\pi/2$ (Kitaev point). While quantum fluctuations eventually remove this degeneracy~\cite{Jackeli2015}, the almost flat precursor of this physics away from the Kitaev point could still be observable.

\vspace*{-0.15cm}
\subsection{Evolution of the spin gap with $K/J$}\label{sec:Gap}
\vspace*{-0.35cm}
The presence of exchange anisotropy and the fact that there is no continuous translational symmetry implies that the crystallization of $\mathbb{Z}_2$ vortices into a superlattice comes with a finite spin gap $\Delta$. 
This is demonstrated in Fig.~\ref{fig:Ed}\,(b), which shows the evolution of $\Delta$ vs $\psi$ for 17 optimal crystals. 
The gap is indeed finite everywhere inside the $\mathbb{Z}_2$VC phase. Its behaviour is non-monotonic and asymmetric with respect to the sign of $K$ (it is significantly larger for $K\!<\!0$ than for $K\!>\!0$).  
The softening of the gap in the vicinity of the C-IC transition ($\psi\!=\!0$) is in accord with the flattening of the ground state energy landscape [see scale in the vertical axis of Fig.~\ref{fig:Ed}\,(a)] and the recovery of the true Goldstone mode at the HAF point.

Of particular interest is the region above $\psi\!=\!0.25\pi$, which shows not only a softening of the spin gap itself [Fig.~\ref{fig:Ed}\,(b)], but also a significant accumulation of low-$\omega$ spectral weight [Fig.~\ref{fig:scattering}\,(a, b)], reflecting the incipient frustrated Kitaev point. Strong quantum fluctuations may thus render this region susceptible to new collective physics that goes beyond our semiclassical analysis, see e.g., the recent study~\cite{Maksimov2019} and \cite{Li2015,Kos2017}.

\vspace*{-0.15cm}
\section{Discussion}\label{sec:Discussion}
\vspace*{-0.35cm}
The prediction~\cite{Rousochatzakis2016} that the coplanar 120$^\circ$ order of triangular Heisenberg antiferromagnets becomes immediately unstable under an infinitesimal Kitaev anisotropy, giving way to {\it mesoscopic} $\mathbb{Z}_2$ vortex crystals, has triggered a significant interest in the community~\cite{Becker2015,Catuneanu2015,Shinjo2016,Trebst2017,Li2015,Kos2017,Kishimoto2018,Maksimov2019}, and remains to be explored and verified experimentally. At present, materials that have been discussed in this context, including the iridate Ba$_3$IrTi$_2$O$_9$~\cite{Dey2012,LeeDo2017,Becker2015,Catuneanu2015}, the mixed-valence iridate Ba$_3$InIr$_2$O$_9$~\cite{Dey2017}, and the rare-earth compound YbMgGaO$_4$ (YMGO)~\cite{YMGO2015a,YMGO2015b,Maksimov2019}, suffer either from intrinsic disorder and impurities or additional complex anisotropic interactions~\cite{Trebst2017}.

The $\mathbb{Z}_2$ vortex crystals can be detected by small-angle neutron or x-ray scattering methods, in analogy to 1D soliton lattices in modulated antiferromagnets (such as Ba$_2$CuGe$_2$O$_7$~\cite{Zheludev1997}) or skyrmion lattices in chiral ferromagnetic helimagnets (such as MnSi\cite{muehlbauer2009} or Cu$_2$OSeO$_3$\cite{Langner2014}). Furthermore, the strongly inhomogeneous magnetization profile near the defected cores of the $\mathbb{Z}_2$ vortices would give rise to characteristic static hyperfine field distributions, which could be probed by NMR or $\mu$SR. 

In this work, we have demonstrated that the $\mathbb{Z}_2$ vortex crystals can also be diagnosed in dynamical spectroscopic experiments in a more direct way. 
We have shown, in particular, that the collective spin dynamics of $\mathbb{Z}_2$ vortex crystals bears two of their most characteristic properties, the large vortex size and the nonlinear, particle-like nature of their defected cores. These show up with a characteristic multi-fragmented intensity profile at intermediate and high frequencies and a set of intense, fully fledged phonon-like modes at low frequencies. 
While certain aspects will be modified in higher orders of the $1/S$ expansion (for example, 
the characteristic high-frequency intensity profile will be further modified by the effect of the magnon decays which are known to be present for non-colinear magnetic orders~\cite{Chernyshev2009,WinterMaksimov2017}), the main qualitative predictions can be used as `smoking guns' for $\mathbb{Z}_2$ vortex crystals in appropriate materials.

\vspace*{0.1cm}
{\it Acknowledgement} -- We thank A. Chernyshev and P. Maksimov for helpful discussions.  This work was supported by the U.S. Department of Energy, Office of Science, Basic Energy Sciences under Award No. DE-SC0018056. We also acknowledge the support of the Minnesota Supercomputing Institute (MSI) at the University of Minnesota.

\appendix
\begin{center}
{\bf Appendices}
\end{center}
\vspace*{-0.1cm}
In these Appendices we provide auxiliary information and technical details on the magnetic unit cells of the $\mathbb{Z}_2$VC  superstructures (App.~\ref{app:MUCs}), the computation of the linear spin-wave spectra (App.~\ref{app:LSWS}), the computation of the DSF and the INS intensities (App.~\ref{app:DSFINS}), as well as the INS profiles for a series of sixteen $\mathbb{Z}_2$VC's (App.~\ref{app:RepINS}). 


\begin{figure*}[!t]
\includegraphics[width=\textwidth,trim= 0 0 0 0,clip]{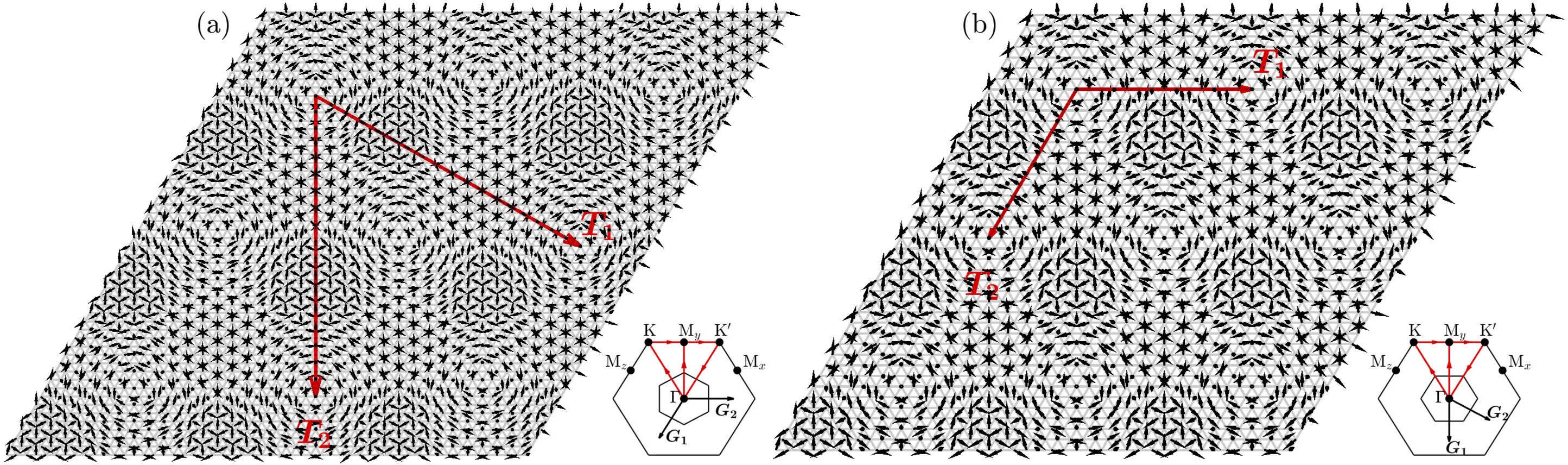}
\caption{Optimal $\mathbb{Z}_2$VC's for (a) $\tan\psi\!=\!-0.31$ ($d\!=\!9$) and (b) $\tan\psi\!=\!-0.34$ ($d\!=\!7$). The hexagons on the right show the first lattice Brillouin zone (BZ, outer hexagon) and the magnetic Brillouin zone (inner hexagon, not in scale).}\label{fig:translation} 
\end{figure*}

\vspace{-0.35cm}
\section{Magnetic unit cells}\label{app:MUCs}
\vspace{-0.35cm}
To perform the semiclassical $1/S$ expansion one needs to deduce the magnetic unit cell for each optimal $\mathbb{Z}_2$VC superstructure. It turns out that the spanning vectors, $\vec{T}_1$ and $\vec{T}_2$, of the magnetic unit cell are of two possible types, depending on the value of $d$ and the sign of the Kitaev interaction $K$. In the first type [Fig.~\ref{fig:translation}(a)], the spanning vectors connect the centers of the domains belonging to the same sublattice, i.e., they connect (A,B,C)$\to$(A,B,C). This type of the magnetic unit cell encloses $\mc{N}_m\!=\!12\,d^2$ spins. In the second type [Fig.~\ref{fig:translation}(b)], which was overlooked in Ref.~\cite{Rousochatzakis2016}, the spanning vectors connect one sublattice to another [(A,B,C)$\to$(B,C,A) for $K\!>\!0$ and (A,B,C)$\to$(C,A,B) for $K\!<\!0$]. This type of magnetic unit cell has $\mc{N}_m\!=\!4\,d^2$ spins. 
The conditions for $d$ and $K$ that give the two different types of magnetic unit cells, along with the associated spanning vectors and number of spins $\mc{N}_m$ in the magnetic unit cell are summarized in Table~\ref{tab:d_req_1}.

\begin{table*}[!t]
\renewcommand{\arraystretch}{1.2}
\begin{tabular}{C{1cm}|C{2.5cm}|C{3.3cm}|C{2.5cm}|C{2.5cm}|C{1cm}}
\toprule
sgn($K$) & $\begin{array}{c}\text{Condition}\\\text{on}~d\end{array}$ & $\begin{array}{c}\text{Sublattice}\\\text{mapping}\end{array}$ & ${\bf T}_1$ & ${\bf T}_2$ &  $\mc{N}_m$\\
\hline
$K\!<\!0$ & $\mod\!(d,3)\!=\!1$ &  (A,B,C)\,$\to$\,(C,A,B) &  $2d\,\vec{a}$ & $2d\,\vec{b}$ & $4\,d^2$ \\  
& otherwise &  (A,B,C)\,$\to$\,(A,B,C) &  $2d\,(\vec{a}-\vec{c})$ & $2d\,(\vec{b}-\vec{c})$ & $12\,d^2$ \\  
\hline
$K\!>\!0$ & $\mod\!(d,3)\!=\!2$ &  (A,B,C)\,$\to$\,(B,C,A) &  $2d\,\vec{a}$ & $2d\,\vec{b}$ & $4\,d^2$ \\  
& otherwise &  (A,B,C)\,$\to$\,(A,B,C) &  $2d\,(\vec{a}-\vec{c})$ & $2d\,(\vec{b}-\vec{c})$ & $12\,d^2$\\
\bottomrule
\end{tabular}
\caption{The two types of magnetic unit cells depending on the period $d$ and the sign of $K$. Spanning vector ${\bf T}_1$ and ${\bf T}_2$, their mapping from one sublattice to another, and number $\mc{N}_m$ of spins in each magnetic unit cell.\label{tab:d_req_1}}
\renewcommand{\arraystretch}{1}
\end{table*}

\vspace{-0.15cm}
\section{Linear Spin Wave (LSW) analysis}\label{app:LSWS}
\vspace{-0.35cm}
In order to study the collective spin dynamics on top of a given optimal $\mathbb {Z}_2$VC, we must first relabel the spin sites $i\to({\bf R},\mu)$, where ${\bf R}=n_1{\bf T}_1+n_2{\bf T}_2$ is the position of the magnetic unit cell ($n_1$ and $n_2$ are integers), and $\mu\!=\!1-\mc{N}_m$ is the sublattice index inside the magnetic unit cell. 
Accordingly, we rewrite the spin ${\bf S}_i$ and its physical position ${\bf r}_i$ as
\be
{\bf S}_i ~\to~{\bf S}_{{\bf R}, \mu}~~~\text{and}~~~
\vec{r}_i =\vec{R}+\bs{\rho}_\mu\,,
\ee
respectively, 
where $\bs{\rho}_\mu$ is the sublattice vector associated to the $\mu$-th sublattice. 
The Hamiltonian is then written as 
\begin{align}
\mc{H}=\frac{1}{2}\sum_{{\bf R}}\sum_{\mu=1}^{\mc{N}_m}\sum_{\nu=1}^{\mc{N}_m} {\bf S}_{{\bf R},\mu}^{\text{T}} \cdot \bs{\mc{J}}_{\mu\nu} \cdot {\bf S}_{{\bf R}+{\bf t}_{\mu\nu}, \nu}\,,
\end{align}
where 
\be
{\bf S}_{{\bf R},\mu}^{\text{T}}=(S_{{\bf R},\mu}^x,S_{{\bf R},\mu}^y,S_{{\bf R},\mu}^z)\,,
\ee 
${\bf t}_{\mu\nu}$ is a primitive translation of the superlattice such that the spins at sites $i=({\bf R},\mu)$ and $j=({\bf R}+{\bf t}_{\mu\nu},\nu)$ interact with each other via $\bs{\mc{J}}_{\mu\nu}$, and 
\be
\bs{\mc{J}}_{\mu\nu}=\left\{\!\!
\begin{array}{rl}
\bs{\mc{J}}_{\bs{\epsilon}}, &  \text{if}~\left(\vec{R}+\bs{\rho}_{\mu}\right)-\left(\vec{R}+{\bf t}_{\mu\nu}+\bs{\rho}_\nu\right) = \pm\bs{\epsilon}, \\ 
	0,  & {\text{otherwise}}
\end{array}
\right.\,,
\ee
where $\bs{\epsilon}\in \{{\bf a}, {\bf b}, {\bf c}\}$ and 
\be
\bs{\mc{J}}_{\vec{a}}\!=\!\begin{pmatrix}
J & 0 & 0\\
0 & {J\!+\!K} & 0\\
0 & 0 & J\\
\end{pmatrix},
\bs{\mc{J}}_{\vec{b}}\!=\!\begin{pmatrix}
J & 0 & 0\\
0 & J & 0\\
0 & 0 & {J\!+\!K}\\
\end{pmatrix},
\bs{\mc{J}}_{\vec{c}}\!=\!\begin{pmatrix}
{J\!+\!K} & 0 & 0\\
0 & J & 0\\
0 & 0 & J\\
\end{pmatrix},
\ee
see Fig.~\ref{fig:lattice}. 
Next, for each site $i=({\bf R},\mu)$, we introduce local reference frames 
\be
\{\widetilde{\vec{x}}_i,\widetilde{\vec{y}}_i,\widetilde{\vec{z}}_i\}
\ee 
such that $\widetilde{\vec{z}}_i$ coincides with the direction of spin ${\bf S}_i$ in the classical ground state. The spin is then rotated into this local frame of reference by a unitary rotation matrix $\vec{U}_\mu$,  
\be
\widetilde{\vec{S}}_{{\bf R},\mu}=\vec{U}_\mu \cdot \vec{S}_{{\bf R},\mu}\,.
\ee 
The matrix $\vec{U}_\mu$ can be constructed using the polar and azimuthal angles $(\theta_\mu,\phi_\mu)$ associated with the direction of the spin in the classical ground state, 
\be
\vec{U}_\mu=\begin{pmatrix}
	\cos\theta_\mu\cos\phi_\mu\quad & \cos\theta_\mu\sin\phi_\mu\quad & -\sin\theta_\mu\\
	-\sin\phi_\mu\quad & \cos\phi_\mu\quad & 0\\
	\sin\theta_\mu\cos\phi_\mu\quad & \sin\theta_\mu\sin\phi_\mu\quad & \cos\theta_\mu\\
\end{pmatrix}.
\ee
Plugging into the Hamiltonian gives
\bea
\mc{H}
=\sum_{\bf R}\sum_{\mu\nu} \widetilde{{\bf S}}_{{\bf R},\mu}^{\text{T}}\cdot\bs{\mc{T}}_{\mu\nu}\cdot\widetilde{{\bf S}}_{{\bf R}+{\bf t}_{\mu\nu}, \nu}\,,
\eea
where $\bs{\mc{T}}_{\mu\nu}=\frac{1}{2}\bs{U}_\mu\bs{\mc{J}}_{\mu\nu}{\bf U}_\nu^{-\!1}$.
Next, we perform a Holstein-Primakoff transformation~\cite{Holstein1940}, and rewrite the spin operators $\widetilde{S}_{{\bf R},\mu}^x$, $\widetilde{S}_{{\bf R},\mu}^y$, and $\widetilde{S}_{{\bf R},\mu}^z$ in terms of bosonic creation and annihilation operators $a_{{\bf R},\mu}^{\dagger}$ and $a_{{\bf R},\mu}$ to lowest order as
\be
\begin{array}{c}
\widetilde{S}_{{\bf R},\mu}^x \!\!\approx\!\!\sqrt{\frac{S}{2}}(a_{{\bf R},\mu}+a_{{\bf R},\mu}^{\dagger})\,,~
\widetilde{S}_{{\bf R},\mu}^y \!\!\approx\!\! -i\sqrt{\frac{S}{2}}(a_{{\bf R},\mu}-a_{{\bf R},\mu}^{\dagger})\,,\\
\\
\widetilde{S}_{{\bf R},\mu}^z =S-a_{{\bf R},\mu}^{\dagger}a_{{\bf R},\mu}\,.
\end{array}
\ee
Then the Hamiltonian can be expanded in powers of $1/\sqrt{S}$, 
\be
\mc{H}=\mc{H}_0+\mc{H}_1+\mc{H}_2+\mc{O}(S^{1/2})\; ,
\ee
where the zeroth-order term
\be
\mc{H}_0=S^2\sum_{\vec{R}}\sum_{\mu\nu}\mc{T}_{\mu\nu}^{(3,3)}
\ee
represents the classical energy $E_{cl}$, the first-order term
\be
\mc{H}_1=\sqrt{\frac{1}{2}}S^{3/2}\sum_{\vec{R}}\sum_{\mu\nu}\Big[\left(\mc{T}_{\mu\nu}^{(1,3)}-i\mc{T}_{\mu\nu}^{(2,3)}\right)a_{{\bf R},\mu}
+ \text{h.c.}
\Big]
\ee
vanishes because we expand around the classical ground state, and the second-order term is 
\bea
\mc{H}_2 &=& 
\frac{S}{2}\sum_{\vec{R}} \sum_{\mu\nu}\Big\{
f_{\mu\nu}  ~a_{\vec{R},\mu}  a_{{\bf R}+{\bf t}_{\mu\nu},\nu} 
+ f_{\mu\nu}^\ast~a_{{\bf R},\mu}^\dagger a_{{\bf R}+{\bf t}_{\mu\nu},\nu}^\dagger  \nn\\
&+&g_{\mu\nu} ~ a_{\vec{R},\mu} a_{{\bf R}+{\bf t}_{\mu\nu},\nu}^\dagger
+g_{\mu\nu}^\ast ~ a_{\vec{R},\mu}^\dagger  a_{{\bf R}+{\bf t}_{\mu\nu},\nu} \nn\\
&-&
2\Big[\mc{T}_{\mu\nu}^{(3,3)}  a_{\vec{R},\mu}^\dagger  a_{{\bf R},\mu}
+\mc{T}_{\mu\nu}^{(3,3)} a_{{\bf R}+{\bf t}_{\mu\nu},\nu}^\dagger a_{{\bf R}+{\bf t}_{\mu\nu},\nu}
\Big]
\Big\}\,,
\eea 
where 
\be
\begin{array}{c}
f_{\mu\nu} = \mc{T}_{\mu\nu}^{(1,1)}-i\mc{T}_{\mu\nu}^{(1,2)}-i\mc{T}_{\mu\nu}^{(2,1)}-\mc{T}_{\mu\nu}^{(2,2)},\\
g_{\mu\nu}  = \mc{T}_{\mu\nu}^{(1,1)}+i\mc{T}_{\mu\nu}^{(1,2)} - i\mc{T}_{\mu\nu}^{(2,1)} + \mc{T}_{\mu\nu}^{(2,2)} \,.
\end{array}
\ee
Using Fourier transform (where ${\bf q}$ belongs to the magnetic BZ) 
\be
a_{{\bf R},\mu}=\frac{1}{\sqrt{\mc{N}_m}}\sum_{\vec{q}} e^{i\vec{q}\cdot(\vec{R}+\bs{\rho}_\mu)} a_{\mu,\vec{q}}\,,
\ee
defining $\bs{\delta}_{\mu\nu}=({\bf R}+\bs{\rho}_\mu)-({\bf R}+{\bf t}_{\mu\nu}+\bs{\rho}_\nu)$, and symmetrizing with respect to $\vec{q}\rightarrow-\vec{q}$, we obtain	
\be
\mc{H}_2 = E_{cl}/S+
\sum_{\bf q} \sum_{\mu\nu}\frac{S}{4} \mc{H}_{2,\vec{q},\mu\nu}
\ee
where
\bea
&&\mc{H}_{2,\vec{q},\mu\nu} =
f_{\mu\nu} \Big[
e^{i\vec{q}\cdot \bs{\delta}_{\mu\nu}}a_{\mu,\vec{q}} a_{\nu,-\vec{q}} 
+e^{-i\vec{q}\cdot\bs{\delta}_{\mu\nu}} a_{\mu,-\vec{q}} a_{\nu,\vec{q}}\Big] +\text{h.c.}
\nn\\
&&
+g_{\mu\nu} \Big[ e^{i\vec{q}\cdot\bs{\delta}_{\mu\nu}} a_{\nu,\vec{q}}^\dagger a_{\mu,\vec{q}}
+ e^{-i\vec{q}\cdot\bs{\delta}_{\mu\nu}} a_{\mu,-\vec{q}} a_{\nu,-\vec{q}}^\dagger\Big] +\text{h.c.}
\nn\\
&&
-2\mc{T}_{\mu\nu}^{(3,3)}\Big[ 
a_{\mu,-\vec{q}} a_{\mu,-\vec{q}}^\dagger
+ a_{\mu,\vec{q}}^\dagger a_{\mu,\vec{q}}
+ a_{\nu,-\vec{q}} a_{\nu,-\vec{q}}^\dagger
+ a_{\nu,\vec{q}}^\dagger a_{\nu,\vec{q}}
\Big],~~~~~~ 
\eea
or in matrix form 
\begin{align}
\mc{H}_2=E_{cl}/S+\sum_{\vec{q}} {\bf x}_\vec{q}^\dagger \cdot {\bf H}_{\vec{q}}\cdot {\bf x}_\vec{q}\; ,
\end{align}
where
$
{\bf x}_\vec{q}=(a_{1,\vec{q}}\, ,...\, ,a_{\mc{N}_m,\vec{q}}\, ,a_{1,-\vec{q}}^\dagger\, ,...\, ,a_{\mc{N}_m,-\vec{q}}^\dagger)^{\text{T}}
$, 
and ${\bf H}_{\vec{q}}$ is a $(2\mc{N}_m)\times(2\mc{N}_m)$ matrix. 
The diagonalization of ${\bf H}_\vec{q}$ involves introducing a new set of Bogoliubov quasiparticle operators~\cite{Bogoliubov1947,Blaizot},
\be
{\bf y}_\vec{q}\!=\!(b_{1,\vec{q}}\, ,...\, ,b_{\mc{N}_m,\vec{q}}\, ,b_{1,-\vec{q}}^\dagger\, ,...\, ,b_{\mc{N}_m,-\vec{q}}^\dagger)^{\text{T}}, 
\ee
obtained from ${\bf x}_\vec{q}$ by a unitary canonical transformation ${\bf x}_\vec{q}=\vec{T_q}\cdot {\bf y}_\vec{q}$. The transformation must be such that the new bosons satisfy the bosonic commutation relation which, in terms of $\vec{T_q}$, gives the condition $\vec{T_q}^\dagger \cdot \vec{g} \cdot \vec{T_q}=\vec{g}$, where $\vec{g}=\text{diag}(\vec{I},-\vec{I})$ and $\vec{I}$ is a $\mc{N}_m\times \mc{N}_m$ unitary matrix. The matrix $\vec{T_q}$ can then be found by solving the eigenvalue equation (in matrix form)\cite{Blaizot}
\begin{align}
(\vec{g}\cdot {\bf H}_\vec{q}) \cdot \vec{T_q}=\vec{T_q}\cdot (\vec{g}\cdot \bs{\Omega}_\vec{q})\; ,
\end{align}
where $\bs{\Omega}_\vec{q}=\vec{T_q}^\dagger {\bf H}_{\vec{q}}\vec{T_q}=\text{diag}(\bs{\omega}_{\vec{q}}, -\bs{\omega}_{\bf q})$, and $\bs{\omega}_{\vec{q}}$ is a diagonal matrix within elements $\{\omega_{1,\vec{q}}, \omega_{2,\vec{q}}, \ldots,\omega_{\mc{N}_m,\vec{q}}\}$.

\vspace{-0.15cm}
\section{Dynamical structure factor (DSF) and inelastic neutron scattering (INS) intensity }\label{app:DSFINS}
\vspace{-0.35cm}
The DSF $S^{\alpha\beta}(\vec{Q},\omega)$ is given by  the Fourier transform of the spin-spin correlations
\bea
\mc{S}^{\alpha\beta}(\vec{Q},\omega)&=&
\sum_{\mu\nu}\int \text{dt}\, e^{-i\omega t}\langle S_\mu^\alpha(-\vec{Q},0)S_\nu^\beta(\vec{Q},t)\rangle \nonumber\\
&=& \sum_{\mu\nu}\int \text{dt} \, e^{-i\omega t}\langle \Big[\frac{1}{\sqrt{\mc{N}_m}}\sum_{\vec{R}}e^{i\vec{Q}\cdot(\vec{R}+\bs{\rho}_\mu)} S_{{\bf R},\mu}^\alpha(0)\Big]
\nn\\
&&\times
\Big[\frac{1}{\sqrt{\mc{N}_m}}\sum_{\vec{R}'}e^{-i\vec{Q}\cdot(\vec{R}'+\bs{\rho}_\nu)} S_{{\bf R}',\nu}^\beta(t)\Big]\rangle\,,
\eea
where the $\alpha$-th component of the spin on the sublattice $\mu$ is given by
\begin{align}
S_{{\bf R},\mu}^\alpha=\sqrt{\frac{S}{2}}\xi_\mu^\alpha a_{{\bf R},\mu}+\sqrt{\frac{S}{2}}\xi_\mu^{\alpha^*} a_{{\bf R},\mu}^\dagger
+\lambda_\mu^\alpha(S-a_{{\bf R},\mu}^\dagger a_{{\bf R},\mu}) ,
\end{align}
and $\xi_\mu^\alpha=[\vec{U}_\mu^{\!-\!1}]^{\alpha,1}-i[\vec{U}_\mu^{\!-\!1}]^{\alpha,2}$, $\lambda_\mu^\alpha=[\vec{U}_\mu^{\!-\!1}]^{\alpha,3}\; .$
Note that the third term in $S_{{\bf R},\mu}^\alpha$ can be dropped when calculating the DSF since this term only describes the reduction of the static ordered moment due to magnon population.

The Fourier transform of the spin component is given by
\begin{align}\label{SQ0}
S_\mu^\alpha(-\vec{Q},0)=&\frac{1}{\sqrt{\mc{N}_m}}\sqrt{\frac{S}{2}} \sum_{\vec{R}} 
e^{i\vec{Q}\cdot(\vec{R}+\bs{\rho}_\mu)} \Big[
\xi_\mu^\alpha a_{{\bf R},\mu} (0)
+\xi_\mu^{\alpha^*} a_{{\bf R},\mu}^\dagger(0)\Big]\\\nn
%
%
%
%
=&\sqrt{\frac{S}{2}} e^{i \bs{\tau}\cdot \bs{\rho}_\mu}  \Big[ 
\xi_\mu^\alpha   a_{\mu,-\vec{k}}(0)
+\xi_\mu^{\alpha^*} a_{\mu,\vec{k}}^\dagger(0)
\Big]\,,
\end{align}
where we used the relation $\vec{Q}=\vec{k}+{\bs{\tau}}$, where ${\bf Q}$ is the momentum transfer, $\vec{k}$ is a wavevector inside the first magnetic BZ, and  ${\bs{\tau}}=n_1\vec{G}_1+n_2\vec{G}_2$ is a primitive vector of the reciprocal lattice of the superstructure, which satisfies $e^{i{\bs{\tau}}\cdot\vec{R}}=1$ for all ${\bf R}$. 
Then the DSF becomes
\begin{align}
\mc{S}^{\alpha\beta}(\vec{Q},\omega)
=&
\frac{S}{2}\int \text{dt} \;e^{-i\omega t}\langle {\bf x}_{\vec{k}}^\dagger(0) \cdot
\left(
\mathbf{V}_\tau^{\alpha^\dagger}\mathbf{V}_\tau^\beta
\right) \cdot
{\bf x}_{\vec{k}}(t)\rangle,
\end{align} 
where $\vec{V}_{\bs{\tau}}^\alpha$ is a vector array of coefficients given by 
\be
\vec{V}_{\bs{\tau}}^\alpha=\left( 
e^{-i{\bs{\tau}}\cdot\vec{r}_1}\xi_1^\alpha,\ldots,e^{-i{\bs{\tau}}\cdot\vec{r}_{\mc{N}_m}}\xi_{\mc{N}_m}^\alpha,e^{-i{\bs{\tau}}\cdot\vec{r}_1}\xi_1^{\alpha^*},\ldots,e^{-i{\bs{\tau}}\cdot\vec{r}_{\mc{N}_m}}\xi_{\mc{N}_m}^{\alpha^*}
\right)\,.
\ee	
Using the Bogoliubov transformation, we then obtain 
\begin{align}
\mc{S}^{\alpha\beta}(\vec{Q},\omega)
=\frac{S}{2}\int \text{dt} \;e^{-i\omega t}\langle {\bf y}_{\vec{k}}^\dagger(0) \cdot \vec{L}_{\vec{k,{\bs\tau}}}^{\alpha\beta} \cdot {\bf y}_{\vec{k}}(t)\rangle \;,	
\end{align}
where the correlation functions of the bosonic quasiparticles are determined by
\bea
&&\langle b_{\gamma,\vec{k}}^\dagger(0)b_{\gamma',\vec{k}'}(t) \rangle=\delta_{\gamma\gamma'}~\delta_{{\bf k}{\bf k}'}~ n(\omega_{\gamma,\vec{k}}) ~e^{-i\omega_{\gamma,\vec{k}} t}\,,\nn\\
&&
\langle b_{\gamma,\vec{k}}(0) b_{\gamma',\vec{k}'}^\dagger(t) \rangle=\delta_{\gamma\gamma'}~\delta_{{\bf k}{\bf k}'}~ \big[ 1+n(\omega_{\gamma,\vec{k}}) \big] ~ e^{i\omega_{\gamma,\vec{k}} t}\; ,
\eea
where $n(\omega_{\gamma,\vec{k}})=[ e^{\hbar \omega_{\gamma,\vec{k}}/(k_BT)}-1 ]^{-1}$ is the Bose factor at temperature T. 
At $T=0$, we therefore end up with
\begin{align}\label{delta}
\mc{S}^{\alpha\beta}(\vec{Q},\omega)
=&\frac{S}{2}\int \text{dt} \;e^{-i\omega t}\sum_{\gamma=1}^{\mc{N}_m} e^{i\omega_{\gamma,\vec{-k}} t} ~\big[ \vec{L}_{\vec{k,{\bs\tau}}}^{\alpha\beta}\big]_{\gamma+\mc{N}_m,\gamma+\mc{N}_m}\nn\\
=& \pi S \sum_{\gamma=1}^{\mc{N}_m} \big[ \vec{L}_{\vec{k,{\bs\tau}}}^{\alpha\beta}\big]_{\gamma+\mc{N}_m,\gamma+\mc{N}_m} \delta(\omega-\omega_{\gamma,\vec{-k}})\; .
\end{align}

Finally, the INS intensity $\mc{I}(\vec{Q},\omega)$ is given by the expression\cite{squires_2012}		
\begin{align}
\mc{I}(\vec{Q},\omega)\propto\sum_{\alpha,\beta}(\delta_{\alpha\beta}-\frac{\vec{Q}^\alpha\vec{Q}^\beta}{\vec{Q}^2})~\mc{S}^{\alpha\beta}(\vec{Q},\omega)\,.
\end{align}

\section{Representative INS profiles}\label{app:RepINS}
\begin{figure*}[!b]
\includegraphics[width=0.97\textwidth]{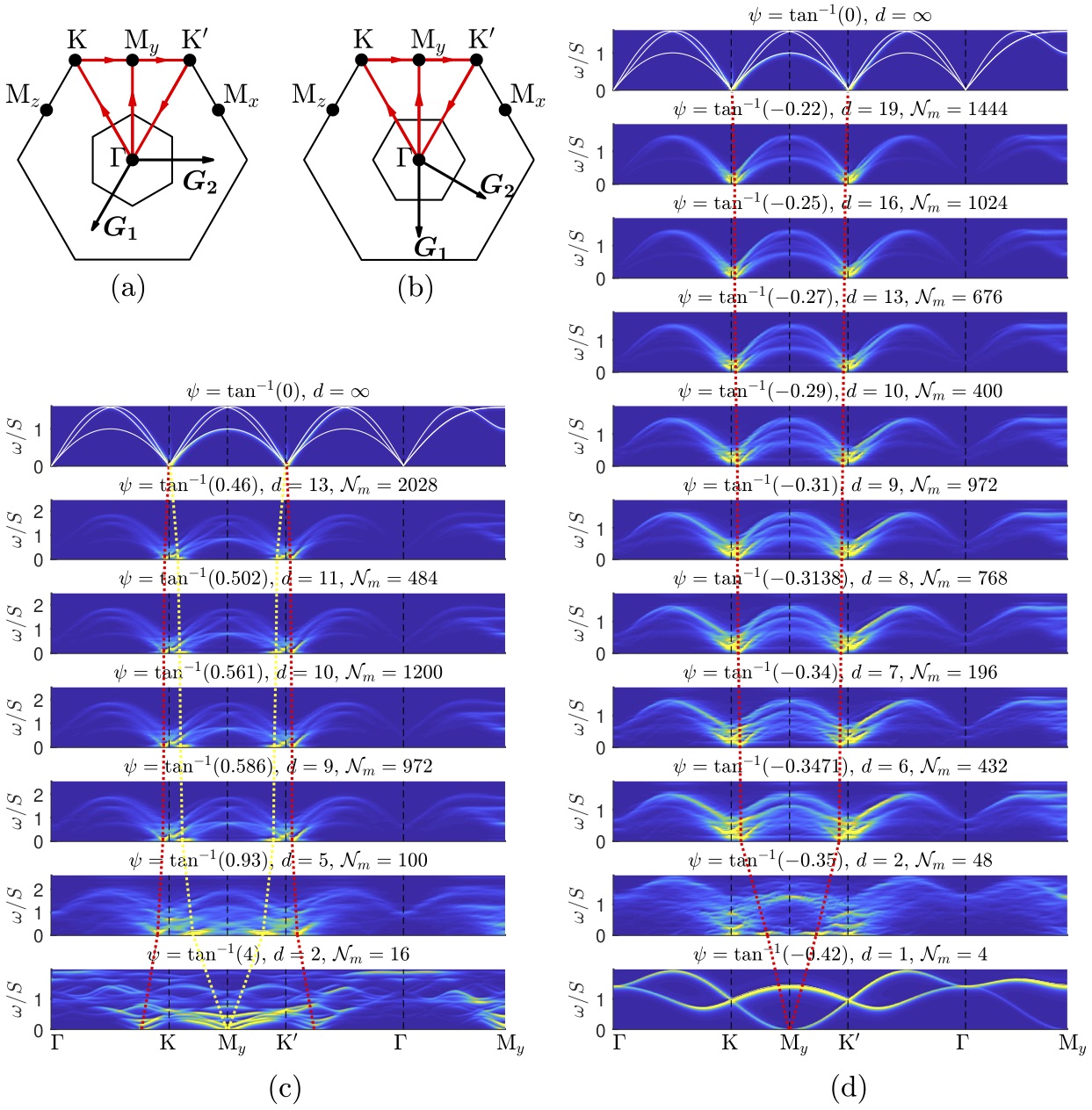}
\caption{ The evolution of INS intensities for various $\mathbb{Z}_2$VC realized at positive (b) and negative (c) values of $\psi$ shown along the high symmetry path shown in (a).} \label{fig:detailedintensity}
\end{figure*}

Figure~\ref{fig:detailedintensity} shows the evolution of the INS intensity $\mc{I}(\vec{Q},\omega)$ for sixteen representative $\mathbb{Z}_2$VC's, as we depart away from the Heisenberg point ($\psi=0$) for both positive (panel c) and negative Kitaev interaction (panel d). The intensity profiles are shown along special symmetry directions in momentum space, see panels (a) and (b). The shift of the positions of the `phonon-like' modes is highlighted by a red dashed curve. These modes follow the positions of the static structure factor. For $K>0$, the positions move from the corner of the BZ $\mathrm{K}$ towards the $\mathrm{\Gamma}$ point, whereas for $K<0$ they move along the directions $\mathrm{K}\rightarrow \mathrm{M}$. The yellow dashed line in panel (c) shows the accumulation of low-$\omega$ spectral weight as we approach the frustrated Kitaev point ($\psi=\pi/2$), see main text.

\clearpage

\bibliographystyle{apsrev4-2}
%
\end{document}